\documentclass[11pt,twoside,letterpaper]{article} %% The same for {book}
\usepackage{times,fancyhdr}
\usepackage[dvips]{graphicx}


\makeatletter 
\def\cleardoublepage{\clearpage\if@twoside \ifodd\c@page\else% 
    \hbox{}% 
    \thispagestyle{empty}%
    \newpage% 
    \if@twocolumn\hbox{}\newpage\fi\fi\fi} 
\makeatother

\def\beq{\begin{equation}}
\def\eeq{\end{equation}}
\def\beqn{\begin{eqnarray}}
\def\eeqn{\end{eqnarray}}

\def\figurename{Figure}
\makeatletter
\renewcommand{\fnum@figure}[1]{\figurename~\thefigure.}
\makeatother

\def\tablename{Table}
\makeatletter
\renewcommand{\fnum@table}[1]{\tablename~\thetable.}
\makeatother

\begin{document}
\title{
\bfseries\scshape Experimental Search for Quantum Gravity}
\author{\bfseries\itshape Sabine Hossenfelder\thanks{E-mail address: hossi@nordita.org}\\
NORDITA, Roslagstullsbacken 23, 106 91 Stockholm, Sweden}
\date{}
\maketitle
\thispagestyle{empty}
\setcounter{page}{1}
% ------- [First Page Running Head] - place it immediately after title! ------
%\thispagestyle{fancy}
%\fancyhead{}
%\fancyhead[L]{In: Classical and Quantum Gravity: Theory, Analysis and Applications \\ 
%Editor: Editor Name, pp. {\thepage-\pageref{lastpage-01}}} % needs \label{lastpage-01} on the last page.
%\fancyhead[R]{ISBN 0000000000  \\
%\copyright~2007 Nova Science Publishers, Inc.}
%\fancyfoot{}
%\renewcommand{\headrulewidth}{0pt}
%------------------------------------------------------------------------------

% ------------ [Running Heads - for odd and even pages] - please insert it only on page 2!
\pagestyle{fancy}
\fancyhead{}
\fancyhead[EC]{Sabine Hossenfelder}
\fancyhead[EL,OR]{\thepage}
\fancyhead[OC]{Experimental Search for Quantum Gravity}
\fancyfoot{}
\renewcommand\headrulewidth{0.5pt} 
%------------------------------------------------------------------------------

\begin{abstract}
We offer a brief survey of existent and planned experimental tests for quantum gravity. First, we outline the
questions we wish to address, and then introduce some of the phenomenological models that are currently used in
quantum gravity, both with and without a lowered Planck scale. After that, we summarize experimental areas 
where these models can be tested or constrained and discuss the status of the field.

This article is partly based on the talks at the workshop on
Experimental Search for Quantum Gravity, Stockholm, July 12-16 2010.
\end{abstract}
\noindent \textbf{PACS}: 04.60.Bc, 11.25.Wx, 11.30.Cp

\vspace{.08in} \noindent \textbf{Keywords:} phenomenology of quantum gravity, Lorentz-invariance violation, space-time foam, emergent gravity, 
deformed special relativity, causal sets, string cosmology, loop quantum cosmology, cosmic strings

\section{The Quest for Quantum Gravity}
\label{intro}

Scientific discovery is like the mapping of unknown territory. Curiosity and the good bookkeeping of scientific inquiry 
gradually revealed to us the ground nature put to our feet, and while some of our knowledge will need revision upon 
closer investigation, today we have access to highly detailed maps and guides of nature, backed up by so much 
evidence we are comfortable to build our lives on this territory. But our curiosity still drives us farther, 
again and again pushing the boundaries of the known, venturing out into the unknown.

There are two ways the expansion of scientific knowledge, the exploration of terra incognita, goes ahead. 
An experiment may test a previously virgin area and make an unexpected find, leaving it for the theorist 
to explain and make sense of. Or, a theorist may put forward a hypothesis and make a prediction, telling 
the experimentalist where to look next. In physics in particular, both has historically gone hand in hand, 
and still does. The theorist aims to make predictions for planned experiments, and the experimentalist 
will be interested in testing well-founded predictions offered by the theorist. Over the centuries, we 
developed methods and procedures that have proven useful in this process and that we rely on today, 
such as peer review and repeatability of experiments. Due to the well-demonstrated 
``unreasonable effectiveness of mathematics in the natural sciences,'' \cite{Wigner} mathematical self-consistency 
of a theory is an essential ingredient assuring quality and success. 

In the following, we will focus on one particular ongoing exploration at the frontiers of our knowledge: The quest for quantum gravity. 

Einstein's theory of general relativity and the standard model of particle physics together provide a remarkably
successful description of our observations. But general relativity has so far refused to be quantized -- it still is an
entirely classical theory. Using the same methods for quantization that were successful for 
all the other interactions in the standard model does not bring the desired result. It's not that gravity cannot be 
quantized -- it is possible to quantize gravity by the standard procedures -- but the result is non-renormalizable and 
can in the best case merely be understood as an effective, not as a meaningful fundamental theory. 

This issue is more than a mere annoyance for the aesthetically inclined theoretical physicist. It signals 
a mathematical tension and a lack of understanding. In other words, it points us into the direction 
of unknown territory where we can expect to gain deep insight into the workings of nature if we 
manage to reveal its secrets. There are three main reasons why the present status requires a solutions:

\begin{enumerate}
\item Quantum particles can exist in superposition states. The best known example is Schr\"odinger's cat which is 
neither dead nor alive, but more realistically a photon traversing a double-slit that is neither here nor 
there. These animals, or particles respectively, carry energy and thus gravitate. Yet we do not know what 
their gravitational field is: as a classical field it does not exist in superpositions. 

\item General relativity predicts the formation of singularities; instances of infinite energy density and 
gravitational forces. 
Such singularities are unphysical and signal a breakdown of the theory. In these extreme areas of spacetime, 
general relativity would have to be replaced with a more fundamental theory. 

\item The black hole information loss problem. Using quantum field theory in a classical black hole 
geometry, Hawking \cite{Hawking:1974sw} showed that black holes emit thermal radiation and thereby lose mass. If 
this radiation remained thermal until the black hole was entirely evaporated, then any distribution 
with the same initial mass that collapsed to a black hole would eventually be converted into the same thermal 
final state. Detailed information contained in the initial configuration would have gotten lost. 
Such an irreversible loss of information however is incompatible with quantum mechanics. One  
hopes that a proper quantization of gravity will solve this contradiction that arises by combining 
quantum field theory with general relativity\footnote{This hope is not without problems either, 
but further discussion would lead us astray. One can also argue that the problem with singularities 
and that of black hole information loss are actually the same problem, see \cite{Hossenfelder:2009xq}.}.
\end{enumerate} 

The two latter arguments are based on weaknesses in our current theories that could be solved by a theory of quantum gravity. 
It is far from clear whether a theory of quantum gravity would solve these puzzles, but it is plausible to expect that it could 
give us a clue how to proceed. The first mentioned problem with superposition states is a stronger argument in that 
its solution necessitates quantum gravity. 

The phenomenology of quantum gravity, which this article is dedicated to, is a still young research field exploring a 
possibility which has previously been neglected in our quest: the possibility to extract knowledge about the 
unknown, looked-for theory from experiment. In the following section, we will survey some of the phenomenological 
models used and, in section \ref{exp}, the experimental possibilities to test them. In section \ref{post}, we will look at the 
possibility that evidence for quantum gravity may be contained in already available data. 

This article is not a review. Rather, it is a introduction into the main concepts 
and ideas of phenomenological quantum gravity. The brief summary provided here, and the 
literature references, are thus necessarily incomplete and I wish to apologize in advance for 
every contribution to the research in this field which did not find a place here. I 
invite the interested reader to send me suggestions for improvement.

\section{Phenomenological Models}

In this section, we will get to know some of the currently
used models for a phenomenology of quantum gravity. The first requirement for a good 
phenomenological model is of course that it be in agreement with already available data 
and that it be internally consistent. Most of the models that are presently used are quite 
young and unresolved issues exist. This makes research in this area lively, and full of 
discussion and controversy. We will however here not go into the details of these discussions,
but take an agnostic point of view in presenting these models as proposals to open the area
of quantum gravity for experimental test.

\subsection{The Planck Scale}

The scale at which effects of quantum gravity are expected to become relevant is the Planck scale, 
named after Max Planck who first introduced these units of energy, length and time in 1899 \cite{Planck}. 
There is an easy way to estimate the Planck scale. Consider concentrating an amount of energy, $E$, 
in a volume of size $\Delta x^3$. Via Einstein's field equations we know that the curvature
of spacetime, a second derivative of the metric, $g$, is related to the energy density. The typical
perturbation that is then caused by the energy is
\beqn
\frac{\delta g}{\Delta x^2} \approx \frac{G E}{c^4 \Delta x^3} \quad, \label{dg}
\eeqn
where $G$ is Newton's constant. We now consider the energy to be localized 
as good as quantum mechanics possibly allows us, i.e. to its Compton wavelength 
$c \Delta x = \hbar/E$. Then one has
\beqn
\delta g \approx \frac{G E^2}{c^3 \hbar} \quad, 
\eeqn
This distortion will become non-negligible when $\delta g \approx 1$,
which happens at a particular energy scale, or mass respectively. In numbers, this mass
scale, which corresponds to the Planck mass, and the related Compton wavelength, the Planck length, are
\begin{eqnarray}
m_{\rm p} = \sqrt{\frac{\hbar}{G c}} \approx 10^{16}{\mathrm{TeV}}~,~ 
l_{\rm p} = \sqrt{\frac{\hbar G}{c^3}} \approx 10^{-20}{\mathrm{fm}} ~,
\end{eqnarray}
and the Planck time is $t_{\rm p} = l_{\rm p}/c$. In the following we will use units with $\hbar = c =1$, 
such that $l_{\rm p} = t_{\rm p} = 1/m_{\rm p}.$ For other intuitive arguments why the Planck
scale marks the regime in which effects of quantum gravity become important, see the nice
article \cite{Adler:2010wf}.

Compared to typical energy scales we are able to reach in our experiments, the Planck
energy is extremely high, too high to hope to be able to test
it directly. This simple fact, a consequence of the gravitational interaction
being so weak compared to the other interactions, is what makes it so difficult to
test quantum gravity. A collider to test quantum gravitational effects would need to 
have the size of our galaxy, and even a detector the size of planet Jupiter wouldn't 
measure as much as a single graviton 
in the lifetime of the universe \cite{shooting7}. This even lead Freeman Dyson to 
hypothesize that the regime in which the standard model of particle physics and general 
relativity lead to conflicting results is entirely undetectable \cite{shooting8}.
However, in the following we will see that the situation is so hopeless not, for 
there are ways to reach the required precision that allows to test some models
of quantum gravity phenomenology.

It is important to note here that our above estimate for the relevance of quantum 
effects relies on a crucial assumption. To obtain the Planck scale, we have extrapolated 
the gravitational interaction from the regime where we have access to experimental 
tests, at around 1 TeV, to about $10^{16}$~TeV. In our simple argument, this has entered through the use
of Eq. (\ref{dg}) which is a consequence of general relativity. But 
that's 16 orders of magnitude in which unexpected new physics can start
to play a role, thereby significantly altering the typical distortion the energy $E$
causes in the space-time geometry. If the gravitational interaction was modified 
at distances smaller than we have tested yet, distances that we have so far not 
been able to resolve with scattering experiments, then our extrapolation might 
fail, and the scale at which effects of quantum gravity become strong can be lower than
what the above estimate suggests.

Scenarios where exactly this happens are models with additional compactified 
spatial dimensions. The relevant feature that these models (discussed in \ref{xd1} and
\ref{xd2}) have in common is that at distances beyond today's experimental reach the 
gravitational interaction is modified through the propagation of gravitons into the 
higher dimensional spacetime. As a result, the Planck scale is lowered and can, 
depending on the parameters of the model, in the most interesting case be accessible 
at the Large Hadron Collider ({\sc LHC}). 

Scenarios with a lowered Planck scale have a distinctively different, 
and easier accessible, signature. We will thus in the following discuss 
them separately.

\subsection{What do we mean with Quantum Gravity?}

Before we continue, let us first clarify what we are interested in, in order to 
outline the topics to consider. It could spring up anywhere, the surprise discovery 
that will turn out to guide us towards a theory of quantum gravity, but for practical 
purposes we will have to restrict our attention to a limited focus area that is 
directly targeted at quantum gravitational effects. 

We will thus leave out here two big area of research that could have potential 
relevance for quantum gravity, that of grand unification and modifications of general 
relativity. The scale at which grand unification is expected to occur is close by 
the Planck scale, and it is a justified hope that a clue to grand unification will 
put us on the right track to a theory also containing quantum gravity, 
commonly called ``Theory of Everything.'' However, we will here exclude all 
searches for physics beyond the standard model that do not explicitly look for 
quantum gravitational effects. Likewise, we will exclude all general searches 
for deviations from general relativity (such as Lovelock or $f(R)$ gravity, 
Horava-gravity, bimetric models, Einstein-Aether gravity, theories of the 
Brans-Dicke type, etc.) –- not because such deviations could not be related to quantum gravity, 
but because we cannot possibly cover all these topics in this limited space, and
they have been covered elsewhere in great detail. A good review on tests of general
relativity can be found in \cite{Will:2005va}, and one on physics beyond the standard
model in \cite{Allanach:2006fy}.

Another area that will not be discussed here are analogue models for quantum gravity,
that are condensed matter systems which mimic features of general relativity and
quantum field theory in a curved background \cite{Barcelo:2005fc}, and
other examples were these similarities exist, such as the recent analogy to Hawking radiation 
using ultra short laser pulse filaments \cite{Belgiorno:2010wn}.
We will also not discuss here condensed matter systems that are dual to some form
of gravity via the conjectured AdS/CFT correspondence \cite{Nastase:2007kj}.
Though these studies may turn out to be useful to better understand quantum gravity,
they do not actually address the Planck scale structure of the space-time that we
live in.

Further, let us be clear that with quantum gravity we do not necessarily mean
a quantization of gravity. If gravity is not a fundamental interaction, but merely 
applicable in a
classical limit, then quantizing gravitational degrees of freedom may not
be the right way to go. Similarly, if quantization is only an emergent
feature, or one approximately valid in the limits that we have tested,
then the quantization of gravity again might not lead us to the right fundamental
theory. Therefore, with quantum gravity we will mean any approach that is able to
resolve the apparent tension between general relativity and quantum field
theories, and to address the three problems mentioned in section \ref{intro} 

\subsection{What is a Phenomenological Model?}

Let us then turn towards the type of models that we will be discussing. 
There are two ways to venture forward into the unknown from where we presently are. 

The one way is a ``top-down'' approach, the attempt to develop a new theory 
from first principles. A common problem with this approach is that not 
only does it typically imply also the development of entirely new 
mathematical techniques, the new theory moreover also must be successfully 
connected back to the theories and observations we already have. This 
connection is presently missing for all approaches towards a fundamental 
theory of quantum gravity. Such top-down approaches necessarily put a 
strong emphasis on mathematical consistency since it is, besides 
inspiration, the only guide at hand.

The other way is a ``bottom-up'' approach, starting from the theories 
we already have, trying to extend them with the hope to discover a 
path leading forward. The problem with this approach is that without 
any directive advice there are very many ways to go. Bottom-up approaches 
are typically considered to merely be bridges towards a more fundamental 
theory. The emphasis on mathematical consistency is thus not quite as 
strong since one can expect these approaches to leave unanswered 
questions for the fundamental theory.

Taking clues from top-down approaches and using them as particular 
bottom-up extensions is then a promising middle-way to take that allows us
to arrive at specifically targeted phenomenological models. In the following, we 
will discuss some examples of such models that one could 
call ``top-down inspired bottom-up approaches.'' While not meant to be fundamental, 
so constructed models aim to predict a phenomenology that tests for specific properties 
the fundamental theory might have. These approaches are in most cases inspired by one 
or the other approach towards quantum gravity, but so far rigorous derivations
from a candidate theory of quantum gravity are missing.

\subsection{Examples of Phenomenological Models for Quantum Gravity}

\subsubsection{Violations of Lorentz Invariance}
\label{LIV}

One of the most frequently asked questions about quantum gravity is whether the 
ground state of space-time obeys Lorentz-invariance or whether this symmetry
is in fact weakly broken. If Lorentz-invariance was broken, observer-independence 
would be explicitly violated and a preferred frame would be singled out. 
This frame is typically thought to be related to the restframe of the Cosmic
Microwave Background ({\sc CMB}) -- a naturally existent and experimentally
confirmed preferred frame of our universe.

Lorentz-invariance violation ({\sc LIV}) has attracted a significant amount of
interest because it appears in many approaches to quantum gravity, though derivations
are not conclusive. There exist Lorentz-violating models inspired by string theory
\cite{Kostelecky:1988zi}, and
it had also been claimed
\cite{Gambini:1998it} that Lorentz invariance is broken in Loop Quantum Gravity ({\sc LQG}) though
it was later argued that the latter calculations were based on the use of 
unphysical assumptions about the ground state of the theory \cite{Lee}. 
A preferred frame appears in non-commutative geometry \cite{Douglas:2001ba}, and
Lorentz-invariance is also violated in many approaches in which gauge bosons are
emergent \cite{emergent}.

Violations of Lorentz-invariance can be realized in different ways.
The most relevant distinction to draw in this class of models is
whether the breaking of Lorentz-invariance is present in the matter sector
of the standard model, or whether it is restricted to the gravitational
sector. 

A breaking of Lorentz-invariance in the purely gravitational sector
is expressed by coupling a tensor field, or its derivatives respectively, 
to geometric quantities constructed from the metric or the curvature. In the most
widely used scenario, the tensor field is a unit, timelike vector field
whose direction singles out the preferred frame. An example of such a
theory is Einstein-Aether theory \cite{Jacobson:2000xp}. However,  as mentioned previously,
such modifications of general relativity will not be discussed 
here. 

A breaking of Lorentz-invariance in the standard model sector can be characterized 
in an effective field theory limit by studying additional terms to the standard
model Lagrangian breaking the symmetry \cite{Colladay:1996iz}. These
additional terms have pre-factors of powers of the Planck mass and a dimensionless
coefficient that should, naturally, be of order one. If the factor is constrained
significantly below that, it speaks against the presence of quantum gravitational effects. 

The features of the resulting theory depend on exactly which terms are present and
may include birefringence (the speed of photons depending on their polarization), 
modification of standard 
model cross-sections, threshold shifts of particle reactions, and
modified dispersion relations (possibly depending on the type of particle) which
generally take the form
\beqn
E^2 - p^2 = m^2 + f(E,\vec p, m_{\rm p}) \quad. \label{mdr}
\eeqn
Typically, $f$ is a power series in $E/m_{\rm p}$, with the prospects for experimental
tests crucially depending on the presence or absence of the first order term.

The phenomenology
of {\sc LIV} models is vast, which has certainly added to their popularity. It should also
be noted that for local, unitary, and Lorentz-invariant quantum field theories, {\sc CPT} symmetry
follows \cite{CPT}. Testing {\sc CPT}-symmetry thus tests for Lorentz-invariance -- provided the
theory is local and has a unitary evolution. 

A well known problem with {\sc LIV} models is that the new (irrelevant)
higher order terms of mass dimension 5 and higher 
induce by radiative corrections terms to the Lagrangian of dimension
4 or less which are ruled out already if the higher order terms had coefficients
naturally expected from quantum gravity. For these models to be consistent and not
in conflict with experiment already, one thus has 
to either believe in fine-tuning, or some mechanism that protects the symmetry of
the lower dimensional operators.

We will in the following not attempt to summarize all experimental tests that have been
made on LIV, but only the most stringent ones to date on higher order terms. For a more 
comprehensive review, the interested reader is referred to \cite{Mattingly:2005re}.

\subsubsection{Deformations of Special Relativity}

Instead of breaking Lorentz-invariance, it has
also been considered the possibility that special relativity may be modified in the high energy regime 
without singling out a preferred frame. This idea has become known as ``Deformed Special Relativity'' 
({\sc DSR}) \cite{AmelinoCamelia:2000ge}. 
In these models, the modified Lorentz-transformations in momentum space have two invariants: A constant, $c$, 
with dimension speed and the Planck mass, $m_{\rm p}$. 

{\sc DSR} has been motivated by {\sc LQG}, though no rigorous derivation exists to
date. There are however non-rigorous arguments that  {\sc DSR} may emerge from a semiclassical limit 
of quantum gravity theories in the form of an effective field theory with  an energy dependent metric \cite{AmelinoCamelia:2003xp},
or that {\sc DSR} (in form of a $\kappa$-Poincar\'e algebra) may result from a version of path integral 
quantization \cite{KowalskiGlikman:2008fj}. In addition it has been shown that
 in 2+1 dimensional gravity coupled to matter, the gravitational degrees of freedom can be integrated out, leaving 
an effective field theory for the matter which is a quantum field theory on $\kappa$-Minkowski spacetime, realizing
a particular version of {\sc DSR} \cite{Freidel:2003sp}. Recently, it has also been suggested that
{\sc DSR} could arise via Loop Quantum Cosmology (see section \ref{LQC}) \cite{Bojotalk}. 

This class of models has some problems that are so far unresolved, most notably the correct description
of multi-particle states and the formulation in position space. It has also been argued 
\cite{Schutzhold:2003yp} that {\sc DSR}
generically causes non-local effects that are in conflict with the standard model.
This issue is still under debate 
\cite{Jacob:2010vr}.

Consequences of a {\sc DSR}-modification of special relativity are that, in most but not all
models, the dispersion relation is modified and the speed of light becomes energy-dependent. The 
constant $c$ that is an invariant of the transformation is then the speed of light in the 
low-energy limit\footnote{It can also be argued that, since the Planck mass as a coupling 
constant should actually be energy dependent too, the second invariant constant of the 
transformations is similarly the Planck mass in the low-energy limit \cite{Calmet:2010tx}.}. 
These features are shared with {\sc LIV} models, but since it has been argued that {\sc DSR}
may not be expressible in form of an effective field theory and to date no (agreed upon)
Lagrangian formulation for the interaction picture exists, most of the constraints on
{\sc LIV} do not apply for {\sc DSR}. It had originally been claimed that {\sc DSR} leads to a shift
of thresholds of particle reactions \cite{AmelinoCamelia:2002gv}, but this claim has later 
been revised \cite{ReviewGAC}.

\subsubsection{Causal Sets} 

The causal sets approach \cite{Sorkin:2003bx} considers as fundamental the causal structure
of spacetime, as realized by a partially ordered, locally finite set of points. This set,
represented by a discrete sprinkling of points, replaces the smooth background manifold of 
general relativity. The ``Hauptvermutung'' (main conjecture) of the causal sets approach is that 
a causal set uniquely determines the macroscopic (coarse-grained) space-time manifold. In
full generality, this conjecture is so far unproven, though it has been proven in
a limiting case \cite{Bombelli:1989mu}.

Intriguingly, the causal sets approach to a discrete space-time can preserve Lorentz-invariance. 
This can be achieved by using not a regular but a random sprinkling of points. It has been shown
in \cite{Bombelli:2006nm}, that a Poisson process fulfills the desired
property. 

This discretization of the background manifold leads to a Lorentz-invariant 
diffusion of energy-momentum; its most general form contains
two phenomenological parameters that can be bounded by the {\sc CMB} spectrum \cite{Dowker:2003hb}.

\subsubsection{Minimal Length and Generalized Uncertainty}

The idea that the Planck length might play the role of a fundamentally minimal
length in nature goes back to Heisenberg in 1938 \cite{Heisenberg}. And indeed, there
are many indications that this is the case (for reviews see \cite{MLreview1}). {\sc LQG} has a minimal
unit of area \cite{MLloop}, in string theory it's string excitations that prevent an arbitrarily
good resolution of space-time structures \cite{MLstring}, and non-commutative geometries \cite{Douglas:2001ba} have
a minimal length scale built in already which results in a finite localization of
wave-packets \cite{MLnoncom}. There are furthermore several thought experiments for probing
smallest distances with test particles that lead one to conclude the non-negligible
perturbations of the background geometry close by the Planck scale result in the
Planck scale setting a limit to how well we can test 
structures \cite{MLthought1,Calmet:2010tx}.
This is not
so surprising because one would expect the Planck length to play the role
of a regulator in the ultraviolet. 

One can then build a model that incorporates the notion of a minimal
length into quantum mechanics and quantum field theory \cite{MLqft}. Features
of these models are a generalized uncertainty principle that prevents
an arbitrarily good localization in position space, a modified dispersion
relation, and a modified measure in momentum space which can be understood
as a non-trivial geometry of momentum space. Such models may or may not
have an energy-dependent speed of light \cite{Hossenfelder:2005ed}. They either break Lorentz-invariance
explicitly or exhibit features akin the previously discussed {\sc DSR},
albeit with a possibly different interpretation \cite{Hossenfelder:2006cw}. The models 
that break Lorentz-invariance or represent a version of {\sc DSR} can be tested via these properties. 
Those that don't typically lead to
predictions, such as modifications of quantum mechanics and cross-sections,
that are far outside possible experimental tests (unless the Planck scale is lowered in which case
they become observable in the same regime as other effects of models
with a lowered Planck scale \cite{Hossenfelder:2004gj}). The most reasonable
place to expect an observable phenomenology is from the high temperature phase in
the early universe. Thus, the thermodynamical properties of these models have
recently been a subject of increased attention \cite{Camacho:2007qy}, though there 
are no predictions yet. 

It is important to note that one does not
need to have a discrete spacetime to have a fundamentally finite resolution of
that spacetime and thus, by employing these models, one does not subscribe to
a particular space-time picture.

\subsubsection{Space-time foam and granularity}
\label{foam}

A central mystery of quantum gravity is the structure of space-time on Planckian
scales. While the details might not be known, the expectation that it is subject
to quantum fluctuations that on shortest distances significantly distort the
background which on long distances seems so smooth, has become known as ``space-time
foam.'' In such models, the background itself is as usual described by a metric. 
There are then different ways to model such fluctuations of the background,
depending on the type of deviations from flat space that one considers. 

One consequence of such models is that the departures of the background geometry from flat 
space on short distances can cause quantum mechanically pure states to evolve into mixed
states \cite{foam}.  This quantum gravitationally caused decoherence,
effectively incorporated in a non-unitary quantum mechanical time-evolution 
leads to violations of {\sc CPT} invariance. Another feature
are fluctuations of light-cones, leading to a stochastic rather than a
systematic departure from a constant speed of light.

An entirely different approach to think about space-time on
Planckian scales is that of a granular structure \cite{Bonder:2008et} which
respects Lorentz-invariance. In this model, the effect of the space-time
structure is argued to appear in a non-trivial coupling between the
Weyl-tensor and the standard model sector.

\subsubsection{Geometrogenesis}

If space-time is fundamentally discrete and described by a network, we find this
network today in a very orderly and highly special state that is well approximated
by a smooth manifold. But one would expect the orderliness of this state to depend
on the temperature, with the network being in a completely different state at high temperatures
in the early universe, and a manifold-like structure only emerging in a phase transition. 

This idea of geometry emerging from a non-geometrical phase has been dubbed ``geo\-metrogenesis'' \cite{Markopoulou:2007jf}, 
and it has been studied in a model of Quantum Graphity \cite{Konopka:2006hu}. In this model, 
one assigns a Hamiltonian to the network which contains information about the node- and link-structure,
and tries to find the ground state and the temperature dependence of the network's regularities. 
This model is still work in progress, but one can hope that in the soon future it will lead
to distinct predictions in the areas of cosmology and astrophysics. 

It has been shown in \cite{Magueijo:2006fu} that a phase transition from a holographic
pre-geometric phase at high temperatures is, under quite general 
conditions, compatible with present day observation. 

\subsubsection{Loop Quantum Cosmology}
\label{LQC}

Loop Quantum Cosmology ({\sc LQC}) \cite{LivRev}  incorporates implications for
the quantum geometry derived from {\sc LQG} in cosmological
settings such that, for sufficiently general parameterizations of the expected
effects, the freedom for instance in quantization choices is
restricted. {\sc LQC} is basically a symmetry-reduced version of {\sc LQG} and in 
this way provides a way to test the consistency of the full theory.  
Currently, cosmological applications are at a
conceptual level, such as suggestions to resolve the singularity
problem, but derivations are moving closer to making contact with
potential observations.

In the context of the singularity problem, several mechanisms have
been suggested, all related to changes in the quantum
space-time geometry (rather than modifications of matter terms that
might violate energy conditions). In isotropic {\sc LQC} the analog of
the Wheeler--DeWitt equation is a
difference equation, which allows
one to evolve the wave-function through geometrical configurations
that correspond to the classical singularity \cite{Sing}. Several models 
(especially those whose matter energy
is dominated by the kinetic term of a scalar field) exhibit
bouncing solutions for their wave-functions \cite{QuantumBigBang}. Once
inhomogeneities can be evolved consistently through the bounce of a
universe model, one
may expect interesting phenomenological implications. However, at present the 
required perturbation theory
is not sufficiently developed to consider inhomogeneities in regimes of
strong quantum corrections that are required to trigger a bounce.

For phenomenology, linear perturbations around spatially flat
Friedmann--Robertson--Walker models have been considered for small
quantum corrections, and cosmological perturbation equations are now available
without having to use gauge-fixing of space-time diffeomorphisms
\cite{ScalarGaugeInv}, which allows one to draw conclusions about the
structure of space-time underlying {\sc LQG}. Interestingly, it turns out
that the quantum corrections cannot merely come
from higher-curvature terms of an effective action, because not just
the dynamics but also the underlying gauge algebra changes. This
deformed gauge algebra  can be
interpreted as a realization of {\sc DSR} \cite{Bojotalk}. The impossibility to
express the corrections as higher-curvature terms means that
quantum corrections can show signatures
characteristic of the quantum-geometry corrections of {\sc LQC}. 
Phenomenological implications are presently being explored mainly for
tensor modes \cite{SuperInflTensor} but also
the more complicated scalar modes are getting under better
control. 

\subsubsection{String Cosmology}

String cosmology is a class of models for the early universe  
that incorporate key principles of superstring theory, such as dilatonic (and axionic) forces,
duality symmetries, winding modes, limiting sizes and curvatures, and higher-dimensional interactions
among elementary extended objects.  The hope is to clarify or resolve some big problems of standard
and inflationary cosmology such as the space-time singularity, the physics of the trans-Planckian
regime, or the initial conditions for inflation.

There are at present four different, but related, string cosmology scenarios \cite{M1}: the pre-big-bang
scenario \cite{M2} (for a review see \cite{M3}), the string gas/brane gas scenario \cite{M4} (for a review see \cite{M5}), the
ekpyrotic/cyclic scenario \cite{M6}, and the brane-antibrane inflationary scenario \cite{M7}. 

In the first three cases 
the initial singularity is removed, and the standard cosmological phase is preceded in time by a 
new epoch dominated by string effects. The standard phase is complemented by an inflationary
but low-energy phase, infinitely extended in time, and asymptotically evolving from the string
perturbative vacuum. In the string gas scenario, the standard cosmological phase emerges from an
initial, higher-dimensional and topologically non-trivial regime in which winding modes and
momentum modes of string and branes are in dynamical equilibrium, thus preventing the
cosmological expansion. 
A nice feature of string gas cosmology is that the gas of winding modes stabilizes 
the size \cite{B7} and shape \cite{B8} of the extra spatial dimensions. 
In the ekpyrotic scenario, an extra spatial dimension is initially shrinking,
and the beginning of the standard cosmological phase is triggered by the collision and the bounce
of the two domain walls representing the space-time boundaries. The absence of a singularity in these
three scenarios stands in contrast to the scenario of brane-antibrane inflation, where the big bang
singularity remains, and where inflation has the conventional dynamics as in standard slow-roll
models. 

These four scenarios do not necessarily exclude each other. For instance, the initial epoch typical of pre-big-bang 
models leads the background cosmological fields towards a regime of strong coupling, and
could be responsible for the process of brane/antibrane production, thus setting the initial conditions
for the phase of brane-antibrane inflation, or for the ekpyrotic bounce, or for the epoch of brane gas
equilibrium. 

For what the phenomenology is concerned, the scenario of brane-antibrane 
inflation leads to basically the same predictions as in
conventional models of slow-roll inflation. 
String gas cosmology provides an alternative to cosmological 
inflation for explaining the origin of structure in the universe. Here, thermal fluctuations 
of strings in the 
Hagedorn phase lead to curvature perturbations which have a scale-invariant spectrum 
like that predicted in inflationary cosmology and the amplitude of the curvature 
fluctuations at late times is in good agreement with the observed value. These
scenarios have further implications for cosmology that will be discussed in
section \ref{cosm}

\subsubsection{Lowered Planck scale: The Arkani-Hamed-Dimopoulos-Dvali Model}
\label{xd1}

In the Arkani-Hamed-Dimopoulous-Dvali (ADD) model \cite{ADD} (for a review, see \cite{ADDreview}), our usual 3+1 dimensional space-time 
has $d$ additional space-like dimensions, each of which is compactified on a radius of 
dimension $R$. All particles that carry standard model charges live on a 3-dimensional 
submanifold (the standard model brane) in that higher dimensional space, whereas gravitons are allowed to propagate 
into all dimensions (the bulk)\footnote{A right-handed neutrino, should it exist, would also
not be bound to the brane.}. This setting is motivated by string theory, in which gravitons are 
described by closed strings that are free to propagate, whereas particles with standard 
model charges are described by closed strings with ends attached to the brane. 

This model explains in geometrical terms why gravity is so much weaker than the other 
interactions, a puzzle known as the ``hierarchy problem.'' Because it is able to propagate into more dimensions, gravity 
dilutes much  faster on short distances ($\ll R$) than as if it was bound to the brane, and on long distances ($\gg R$), 
where we recover to good accuracy 3+1 dimensions,
gravity thus appears weakened relative to the other interaction. 
The relation between the fundamental, higher dimensional Planck scale, $M_{\rm f}$, 
and the Planck scale we have measured at long distances is given by the volume of the 
extra dimensions
\beqn
m_{\rm p}^2 = R^{d} M_{\rm f}^{d+2} \quad.
\eeqn

If one inserts some numbers, assuming the fundamental scale $M_{\rm f}$ is around a TeV, one 
finds that for $d>2$ the radius of the extra dimensions is far below the distances for which we have direct tests of
the gravitational potential, with the case $d=2$ meanwhile being tightly
constrained by precision measurements \cite{Masuda:2009vu} (and the case $d=1$ is ruled out -- the
extra dimension would need to have the size of the solar system). Note however that the 
{\sc ADD}-scenario does not actually solve the hierarchy problem, but merely 
reformulates it, since the inverse of the radius $R$ has to be many orders of magnitude 
smaller than $M_{\rm f}$, which introduces a new unexplained hierarchy.

The most important consequences of this scenario are that on short distances,
gravity is stronger than without extra dimensions, and that gravitons can
have a non-vanishing momentum-component into the direction of the extra dimensions.
The graviton's momentum is then geometrically quantized in multiples of the inverse
radius. On the brane, these graviton excitations -- also referred to as ``the
Kaluza-Klein-tower'' -- appear like massive gravitons.

The {\sc ADD} model has attracted a lot of attention during the last decade and 
exists in many variants with different radii of extra dimensions, 
compactifications on different topologies, or taking into account a finite 
width of the brane (which itself could have an internal structure), brane recoil,
and many other details that we will not go into here.

\subsubsection{Lowered Planck scale: The Randall-Sundrum Model}
\label{xd2}

The Randall-Sundrum (RS)-model \cite{RS} (for a review, see \cite{RSreview}) equips space-time with one 
additional space-like  dimension and, as in the {\sc ADD}-model, gravitons are allowed to propagate
into the full space-time, while particles with standard model charges are
bound to a 3+1 dimensional submanifold. In contrast to the ADD-model however, 
the geometry of the extra dimension in the RS-model is not flat but curved
with an exponential function of the coordinate of the extra-dimension, the so-called ``warp-factor.'' 
In the bulk, one has a negative cosmological constant, and the space-time is thus a slice of 
Anti de-Sitter (AdS)
space with the standard-model-brane, also called the TeV-brane, on one end, and another brane called the Planck-brane
on the other end. In one variant of the model (known as RS2), this slice is infinitely extended, but we
will here focus on the case (RS1) where it is of a finite size $\pi R$.

The relation between the
fundamental scale, $M_{\rm f}$, and the usual Planck scale is in the RS-model given
by
\beqn
m_{\rm p}^2 = \frac{M_{\rm f}^3}{k} \left( 1 - e^{ - 2 \pi k R} \right) ~,
\eeqn
where $k$ is the curvature radius of the AdS slice. To examine the strength of gravity on the TeV-brane, one
considers the standard model Lagrangian on that brane. One finds that one can absorb 
all factors stemming from the non-trivial metric in field redefinitions and thus
reproduce the usual Lagrangian,  except that the vacuum expectation value 
of the Higgs is exponentially suppressed by a factor $\exp{(-\pi k R)}$, thus explaining
the smallness of particle masses relative to the Planck mass. To obtain the observed
value, one then has to
choose the product $R k$ accordingly. By merit of
the exponential function, it is here not necessary to introduce another hierarchy
to achieve a factor of many orders of magnitude -- it is sufficient if $kR$ is
of order ten.

The wave-equations for the gravitons in the background geometry of the RS-model
are Bessel equations. Their solutions lead to a discrete, though not
periodic, mass spectrum for graviton excitations with a spacing of the order $k\exp{(-\pi k R)}$.

As the {\sc ADD}-model, the RS-model meanwhile exists in many variants -- different
field localizations, thick branes, additional dimensions -- that we will not discuss here.

\subsubsection{Other}

An interesting proposal that should be mentioned here whose origin dates 
back more than 30 years \cite{Kibble:1976sj} is that of
cosmic strings (for a review see \cite{CosmicStrings}). Cosmic strings are 1+1 dimensional
defects formed in the early universe that, under appropriate conditions,
can grow during the evolution of the universe. Cosmic strings are
characterized by their energy density. Today, they would form a network
of (infinitely) long strings and closed loops. 

Originally considered to be caused
by the breaking of axial symmetries in quantum field theories, cosmic
strings have recently attained renewed interest as they could appear
also as relics of fundamental strings \cite{cosmicstringstring}. While
cosmic strings of either type would be an important discovery allowing
us a glimpse into the early universe, for the sake of quantum gravity
phenomenology the question is whether it would possible to distinguish
fundamental strings from the quantum field theory strings. 

%General tests of foundations of quantum mechanics 
%(expl whether equivalenc principle holds for quantum matter, higher order 
%interference etc). Carlip, Penrose?

\section{Experimental Search and Predictions}
\label{exp}

In this section, we will summarize the most important experimental
tests of the previously introduced models. Neutrino-physics will be
discussed in a separate section \ref{neutrino} rather than be broken 
down by the various sources of neutrinos.

\subsection{Collider Searches}

The most powerful particle colliders in operation today are the Tevatron
at Fermilab, and the {\sc LHC} at {\sc CERN}. These experiments are allowing us to
peer into the structure of matter on distance scales down to $\approx 10^{-18}$ m.
If the Planck scale is lowered, in either the {\sc ADD}- or the RS-scenario, 
it might become accessible in collider experiments. To solve the Hierarchy 
problem, one expects the ``true'' Planck scale to be not too much above the 
electroweak scale ($\approx 250$ GeV), thus hopes are that the Tevatron has established lower
bounds, and {\sc LHC} may be able to catch a 
glimpse of quantum gravity. 

The relevant signatures in this case are the production of gravitons, modifications 
of standard model cross-sections by virtual graviton exchange, and the production 
of black holes. The production of black holes in particular would be a spectacular 
observation, giving us an experimental handle on the black hole information loss problem. Present Tevatron
data allows to put constraints on the parameters of both the RS- and the {\sc ADD}-model, in particular 
from precision electroweak observables. For the {\sc LHC}, numerous detailed predictions 
have been made, including many subclasses of models and specific scenarios that are presently
being tested. An up-to-date summary of the 
best current constraints on the model parameters in both the RS and the {\sc ADD}-model, including
the relevant references, can be 
found in the Particle Data Book \cite{PDB}. As one can expect, the constraints on the
new fundamental scale are of the order TeV. In the {\sc ADD}-scenario the exact constraints 
are usually dependent on the number of extra-dimensions.

\subsection{Astrophysics} 

Astrophysical processes can occur at higher energies than we are presently able
to achieve in particle colliders, but they happen in a less controlled experimental
environment which implies additional uncertainties. For this reason, astrophysical
and collider constraints often complement each other. 

Next to the previously discussed constraints from collider physics, the ADD-model is
also tightly bounded by astrophysical data. The emission of massive gravitons
can lead to an additional cooling of supernovae \cite{Cullen:1999hc}, radiative decays 
of these gravitons can give rise to a diffuse cosmic gamma-ray background \cite{Hannestad:2001jv}, 
and gravitationally trapped massive gravitons can result in a re-heating of supernovae remnants 
and neutron-stars \cite{Hannestad:2001xi}. These effects lead to bounds on the model's
parameters that are also summarized in the Particle Data Book \cite{PDB} and, for less than 4 extra
dimensions, push the new fundamental scale above the range testable at the {\sc LHC}. There
are moreover constraints from ultra high energetic cosmic rays ({\sc UHECR}), which are
similar to the ones from collider physics -- though the energy is somewhat higher still, 
additional theoretical and experimental uncertainties come in to play. With this,
we will leave now the scenarios with a lowered Planck scale.  

The tightest constraints on Lorentz-invariance violating higher order operators
come today from astrophysics. By analyzing the electromagnetic radiation from
the Crab Nebula, the relevant parameters for order 5 terms in an extension of
quantum electrodynamics \cite{Myers:2003fd} have been constrained
to $\xi< {\cal O}(10^{-7})$ and $| \eta_{\pm} | < {\cal O}(10^{-5})$ \cite{Maccione:2007yc}. 
We remind the reader that the natural value one would expect from quantum gravitational
effects is of order one. 

One of the predictions of {\sc DSR} that is presently tested is that of an energy-dependence
in the arrival time of highly energetic photons from distant $\gamma$-ray-bursts ({\sc GRB}) \cite{AmelinoCamelia:1997gz}. Such
an effect can also arise from {\sc LIV} but, as previously mentioned, {\sc LIV} models
are tightly constrained already, whereas {\sc DSR} evades these constraints and thus
a parameter $\alpha$ of order one for first order deviations in the dispersion relation 
is still under consideration. The time-delay $\Delta T$ one expects between two photons
with an energy difference $\Delta E$ from Eq.(\ref{mdr}) for such 
a first-order deviation 
is 
\beqn
\Delta T = \alpha \frac{\Delta E}{m_{\rm p}} L ~,
\eeqn
where $L$ is the distance the photons have travelled. If one inserts some numbers, one finds
that the delay can be of the order seconds for a distance of some Gpc and some GeV of energy,
if $\alpha$ is of order one. 
The best constraint to date
comes from the GRB090510, which puts the limit at $\alpha < 1.2$ for the case in which
higher energetic photons are slowed down \cite{:2009zq}. It is significantly harder to
put bounds on a scenario in which the effect is stochastic, and better statistic is
needed for that. 

It has also been suggested that a modified dispersion relation as it appears in {\sc LIV} and {\sc DSR} may
be tested with weak gravitational lensing \cite{AmelinoCamelia:1997gz,Biesiada:2007rk}, but the effect is 
out of presently possible precision.

Cosmic strings, should they exist, would cause particular gravitational lensing images
that however would not allow us to distinguish whether these strings are of fundamental
origin. More interesting for our purposes is thus that cosmic strings would generically create cusps 
and be sources of gravitational radiation. This could be detected in the soon future using experiments 
like {\sc LIGO} or {\sc LISA}. In the case of the fundamental cosmic super-strings that are interesting 
for quantum
gravity phenomenology, the presence of extra dimensions enhances the number density of loops
in the strings' network \cite{Damour:2004kw}. It has however been pointed out recently that
taking into account the full dynamics of the string in presence of extra dimensions larger than the 
width of the string also has the effect of lowering 
the probability of cusp formation and of
dampening the gravitational wave emission, thus making the detection more
difficult \cite{O'Callaghan:2010ww}. 
 
Light-cone fluctuations that appear in models of space-time fuzziness can lead to 
spectral line broadening and angular blurring of distant sources \cite{Thompson:2006qe}. This
effect too is unfortunately not detectable with presently possible precision.

\subsection{Cosmology}
\label{cosm}

The best place to look for quantum gravitational effects is in regions of strong
curvature, i.e. towards the center of black holes or the early universe. Since
the black hole interior is hidden from observation by the horizon, this makes
the early universe the prime candidate. 

Our currently most remarkable insights
about this era stem from the {\sc CMB} whose temperature anisotropies carry an imprint of the background 
perturbations back then. The {\sc CMB} radiation is polarized, with the polarization
pattern commonly decomposed into $E$-modes (curl-free) and $B$-modes (divergence-free), 
in an analogy to electrodynamics. The detection of $B$-modes provides a signature of primordial gravitational
waves (tensor modes) and via those contains information about the dynamics of the background. The
precision of $B$-mode measurements is expected to significantly increase in the soon future
through experiments like QUaD, BICEP, ABS, and Quijote. For the following, it is useful to 
know that in the standard (slow-roll, single field)
inflationary scenario the quantum fluctuations of the
metric tensor have a flat or almost flat spectrum, and the {\sc CMB}-fluctuations are to good precision Gaussian.

In the pre-big-bang scenario of string cosmology, the quantum fluctuations of the
metric tensor tend to be amplified with a spectrum which is rapidly growing in frequency. 
This has two important physical consequences. First, the rapidly growing 
spectrum of the tensor metric perturbations leads to a relic background of pre-big-bang gravitons which 
peaks at high frequency \cite{M11}. This signature should be 
easily accessible to planned experiments for Earth-based detectors (e.g.
{\sc LIGO}, {\sc VIRGO}) and gravitational antennas operating in space (e.g. {\sc LISA}, {\sc BBO}, {\sc DECIGO}).
At the low frequency scales relevant to the observed {\sc CMB}-anisotropy, the spectrum is in contrast strongly 
suppressed and a possible contribution to the {\sc CMB}- polarization should be completely negligible \cite{M12}.
Second, the non-standard production of {\sc CMB}-anisotropies in the pre-big-bang scenario induce a 
small non-Gaussianity in the {\sc CMB}-spectrum, which may be detectable in the soon future: with
the {\sc WMAP} 8-year mission results one expects a 20\% improvement on the bounds on non-Gaussianity, 
while ESA's Planck satellite can yield a factor of about 4.

In the ekpyrotic string cosmology scenario the relic background of primordial gravitons also has a  
growing spectrum, but is expected to be quite negligible today both in the high-frequency and low-frequency
range \cite{M13}. In string gas cosmology, the spectrum of gravitational waves is predicted to 
be mostly scale-invariant but also with a slight blue tilt \cite{B6}. The scenario of brane-antibrane inflation 
finally basically leads to the same predictions as in conventional models of slow-roll inflation.

In {\sc LQC}, scenarios without a bounce produce a tensor mode spectrum quite similar to that
in the ekpyrotic string cosmology scenario \cite{SuperInflTensor}. In case of a bounce, the power of tensor modes 
has also been found to be suppressed at small frequencies 
relative to the standard scenario, but with a characteristic bump at an intermediate scale
and approaching the standard result at high frequencies \cite{Mielczarek:2010bh}. The model parameters
of this prediction (the mass of the inflaton field and the position of the peak in the spectrum) 
were recently constrained with presently available data 
on the {\sc CMB} $B$-modes from QuAD and {\sc BICEP} in \cite{Ma:2010yb}. One can expect these
constraints to tighten significantly with better data.

Cosmic strings are yet another mechanism to produce non-Gaussianities \cite{Ringeval:2010ca},
and they also affect the spectrum of {\sc CMB} $B$-modes \cite{Ma:2010yb}, both of which
yields constraints on their abundance.

The Lorentz-invariant diffusion that appears in a model inspired by the causal sets approach also
affects the polarization of {\sc CMB} photons \cite{Contaldi:2010fh}. Presently available
data was found to be consistent with no effect, though with a slight bias for a small
effect.

\subsection{Neutrino Physics}
\label{neutrino}

Neutrinos are interesting for the purposes of testing feeble quantum gravitational
effects for two reasons. First, their very small masses and second, their weak
interaction which enables them even at high energies to travel long,
possibly cosmological, distances almost undisturbed. Neutrino experiments
fall into different categories, depending on the source of neutrinos: Earth
based (reactor, collider, and other man-made neutrino sources), solar neutrinos, 
atmospheric neutrinos, and cosmogenic neutrinos.
From those, the latter reach the highest energies and longest travel times,
exceeding $10^{20}$~eV and some Gpc distance. On the other hand, their
flux is small at high energies, such that collecting useful statistics is difficult.

In \cite{Mattingly:2009jf}, it has been suggested to use cosmogenic neutrinos to tighten bounds on 
{\sc LIV} with upcoming experiments. In \cite{Christian:2004xb} 
it has been pointed out that cosmogenic neutrinos could be used to test 
modified dispersion relations to high precision if a baseline of $L \sim \pi m_{\rm p}^4/E^5$
could be reached, where the flux-ratio of different neutrino species would
become sensitive to the distance. And though the energies and
distances needed are so far not achievable, technological progress in that
area is rapid. Cosmogenic neutrinos might at some point also be able
to constrain scenarios with a gravitationally induced collapse of the 
wave-function \cite{Christian:2005qa}.

In some models, {\sc CPT}-violation can be a cause of neutrino-oscillation 
(rather than a mass-difference) \cite{Kostelecky:2003xn}, though the simplest models are 
already ruled out by data \cite{Barger:2007dc}. Such scenarios have the additional virtue of 
possibly accommodating the {\sc LSND} and MiniBoone data \cite{Hollenberg:2009ak}.
Their challenge is typically to be able to fit all the present data without
introducing too many parameters.

It has further been proposed to combine neutrino measurements
with the detection of photons from $\gamma$-ray bursts to 
better constrain modified dispersion relations in {\sc DSR} and {\sc LIV} \cite{AmelinoCamelia:2009pg}.

\subsection{Other}

Among the effects that do not fall into any of the above categories counts for example 
a particular sort of holographic noise induced by space-time fluctuations that
may be detectable in presently operating gravitational wave interferometers \cite{Hogan:2009mm}. 
The model of space-time granularity (\ref{foam}) is testable in an E\"otv\"os-like 
experiment \cite{Bonder:2010qk} and might soon be subject to constraints. Finally,
it should be mentioned the neutral Kaon doublet whose oscillations allow for tests
of {\sc LIV} and {\sc CPT}-violation induced by decoherence. The
precision of these tests for decoherence is presently at the level where one expects
effects of quantum gravity to become important if the relevant parameters have
natural values of order one \cite{Adler:1995xy}. 
%Quantum gravitationally induced collapse of the wave-function

\section{Postdictions}
\label{post}

In the previous section, we have summarized so far unexplored areas in which we might be 
able to find signal of quantum gravity, at higher energies, better precision, larger 
distances or combinations thereof. But there is also the possibility that we have already seen signatures of 
quantum gravity, yet not recognized them for what they are for we have today experimental 
data which is insufficiently understood, and thus points at shortcomings in our current theories.  

The most obvious shortcoming in our ability to explain available data we are facing in cosmology. 
An increasing amount of cosmological and astrophysical observations with ever better precision
show that 96\% of the universe's content has peculiar properties, different to those of matter we 
are made of. Approximately 74\% is ``dark energy'' and 22\% is ``dark matter'' \cite{WMAP}. Dark energy obeys
an unusual equation of state that has the consequence of the universe's expansion accelerating.
The equation of state of dark matter is that of our usual matter but, as dark energy, it is
non-luminous and thus, so far, evading direct detection. While we can describe the behavior
of dark energy and dark matter on a macroscopic level for the purposes of cosmology and 
astrophysics, we do not presently know their microscopic origin. Proposals for both are
abundant. 

Other unexplained data is rather banally the masses of 
elementary particles. Why they have the values we observe nobody knows. Likewise nobody 
knows why the particles classify in three generations, or why we live in 3+1 dimensions.

It might be that some of these puzzles find their answer in quantum gravity. 
Just to mention a few ideas in this area: The cosmological constant for example might be a result 
of a specific sort of defects in a discrete space-time slicing, which are non-local links 
connecting points that have a macroscopic distance in the background manifold \cite{PrescodWeinstein:2009wq}. 
In \cite{Alexander:2009yb} it was put forward a proposal after which neutrino-masses and 
the cosmological constant are emergent, with additional predictions of Lorentz- and 
{\sc CPT}-violating neutrino oscillations. The cosmological constant has also a natural
place in the causal sets approach \cite{Sorkin:1997gi}. It has further been suggested that
a string theory inspired model might be an alternative to dark matter when it comes
to galaxy rotation curves \cite{Cheung:2007tt}, and string gas cosmology 
provides a mechanism for explaining why only three of the nine spatial
dimensions which string theory contains can become macroscopic \cite{B1}. 

We will certainly learn more about the viability of these proposals with 
running or upcoming direct searches for dark matter and more precise
tests for the equation of state of dark energy, for example from high-redshift 
supernovae.

\section{Summary}

The phenomenology of quantum gravity is a young and lively research field that brings
together theoretical and experimental physicists from many areas. As such, it
faces particular challenges. One is that the predictions of different models
may be very similar. To address this issue, the various predictions have to be contrasting 
directly, possibly combining several experimental areas. This has been done very well with the 
{\sc LHC} predictions from extra dimensional models, but similar efforts are presently 
lacking for example when it comes to different scenarios for the early universe. The other 
challenge is in the collaboration and communication between theory and experiment, an exchange 
that becomes ever more important the more involved data analysis becomes. 

To date we have no experimental signature for quantum gravitational effects. But,
as we have seen in the previous sections, creativity, persistence and technological
improvements have brought us closer to this goal. The quest for quantum gravity
may proceed slowly and sometimes be frustrating, but the reward will be nothing less than 
a revolution of our understanding of space and time.

\section*{Acknowledgements}

I thank Ed Copeland, Ruth Gregory and Stefan Scherer for clarifying discussions, and I am especially grateful to Robert Brandenberger, Martin Bojowald, Maurizio Gasperini
and Lee Smolin for their helpful contributions to the work on this manuscript. I further thank all participants of the workshop on Experimental Search for Quantum Gravity that took place at {\sc NORDITA}, 
Stockholm, July 12-16, 2010. The interested reader can find recordings of
the talks at {\tt www.nordita.org/esqg2010}.

\small{
}

\end{document}